\documentclass[12pt]{article}

\usepackage{siunitx}
\usepackage{optidef}
\usepackage{tabularx}
\usepackage[numbers]{natbib}
\usepackage{bm}
\usepackage[normalem]{ulem}

\DeclarePairedDelimiter{\inner}{\langle}{\rangle}

\begin{document}

\title{Optimal Sampling Design Under Logistical Constraints with Mixed Integer Programming}
\author{Connie Okasaki$^{1,2}$\thanks{This material is based upon work supported by the National Science Foundation Graduate Research Fellowship under Grant No. DGE-1762114.},
S\'{a}ndor F. T\'{o}th$^{1,3}$,
Andrew M. Berdahl$^{1,2}$\thanks{AMB was supported by the H. Mason Keeler Endowed Professorship in Sports Fisheries Management.} \\~\\
$^{1}$Quantitative Ecology and Resource Management Program, \\
University of Washington (UW) \\
$^{2}$School of Aquatic and Fisheries Science, UW \\
$^{3}$School of Environmental and Forestry Science, UW
}

\maketitle




\begin{abstract}

The goal of survey design is often to minimize the errors associated with inference: the total of bias and variance. Random surveys are common because they allow the use of theoretically unbiased estimators. In practice however, such design-based approaches are often unable to account for logistical or budgetary constraints. Thus, they may result in samples that are logistically inefficient, or infeasible to implement. Various balancing and optimal sampling techniques have been proposed to improve the statistical efficiency of such designs, but few models have attempted to explicitly incorporate logistical and financial constraints. We introduce a mixed integer linear program (MILP) for optimal sampling design, capable of capturing a variety of constraints and a wide class of Bayesian regression models. We demonstrate the use of our model on three spatial sampling problems of increasing complexity, including the real logistics of the US Forest Service Forest Inventory and Analysis survey of Tanana, Alaska. Our methodological contribution to survey design is significant because the proposed modeling framework  makes it possible to generate high-quality sampling designs and inferences while satisfying practical constraints defined by the user. The technical novelty of the method is the explicit integration of Bayesian statistical models in combinatorial optimization. This integration might allow a paradigm shift in spatial sampling under constrained budgets or logistics. 

\end{abstract}

\section{Introduction}
Since the early 1900s, random sampling methods have been preferred for their unbiased estimators, derived from the controlled introduction of randomness through the design process \citep{rao1990history}. However, these unbiased estimators rely on the assumption that any random sample generated by the design process would in fact be possible to collect. Under complex logistical and budgetary constraints, this assumption may not always hold. Some methods, such as cluster sampling, have arisen to handle particular types of logistical constraints \citep{lohr2019sampling}, but we are not aware of a randomized method capable of flexibly guaranteeing satisfaction of a wide range of constraints. Practical, consistent, and rigorous implementation of complex surveys subject to budgetary constraints can therefore require a great deal of effort, and sometimes require making compromises between these factors \citep{hughes2008acquiring}. In large-scale spatial surveys in particular, heterogenous costs, access to sampling sites, transportation of crews and samples, and other factors can complicate the implementation of a design. In this chapter, we use spatial sampling as an example to demonstrate and quantify the performance of our novel method for conducting a model-based optimal sampling design, in which logistics are explicitly modeled and feasibility of the resulting design is guaranteed.

Apart from complex logistical and budgetary constraints, spatial sampling poses a second major challenge, spatial autocorrelation, which can be summarized by the proverbial Tobler's Law that ``everything is related to everything else, but near things are more related than distant things'' \citep{tobler1970computer}. In the absence of independent samples, the quality of inference is therefore dependent upon the entire set of sampled locations, rather than a summary statistic such as the sample size. This adds an additional layer of complexity. The premise of our model is that combinatorial optimization can serve as a mathematical frame within which statistical models as well as the practical constraints of sampling can be captured simultaneously and explicitly, thereby providing high-quality designs that are guaranteed to be feasible on the ground.  

In particular, we propose to use a $C$-criterion as the objective function of a discrete mathematical program. This criterion can be expressed as $\bm{v}^T\bm{M}^{-1}\bm{v}$ for a fixed coefficient vector $\bm{v}$ and a matrix $\bm{M}$ which is optimized by the model. In the proposed application of spatial sampling, the vector $\bm{v}$ can be chosen in such a way that this criterion measures uncertainty for an estimate of an areal mean, which is a common target of inference for randomized designs \citep{wang2012review}. 
Furthermore, because we express our method as a MILP, logistics can be modeled with a great deal of complexity and sophistication. MILP models are a well-developed branch of operations research, and have for decades been used to model a wide range of industrial applications \citep{dantzig1957discrete}. 

\subsection{Motivation}

Our method is motivated by the following case study. The United States Forest Service (USFS) conducts a regular Forest Inventory and Analysis (FIA) program across the United States, in which they report forest area, tree growth and mortality, and land ownership, among many other metrics. Sampling forests in remote regions such as interior Alaska, however, poses a logistical challenge. The USFS has developed a complex two-phase design and post-stratified estimation procedure for these regions (Chapter 4, \cite{bechtold2005enhanced}). Although the resulting data are flexible and may be used for a wide variety of purposes, the design has become increasingly difficult to implement as more remote regions need to be sampled. This challenge has motivated our development of an alternative, model-based method of design that can guarantee feasibility. 

Our model can also be applied in a broad range of other scenarios. MILPs can be used to generate detailed optimal routes for vehicles such as remote sensing aircraft (e.g. \cite{avellar2015multi}), or fleets of autonomous aquatic robots (e.g. \cite{yilmaz2008path}), or to plan the placement of stationary sensors (e.g. \cite{kwak2022sensor}) or watercraft inspection sites (e.g. \cite{fischer2021managing}). Under these and other sufficiently complex logistics, randomized samples may be difficult to practically collect (e.g. \cite{hughes2008acquiring}), and even if the sample is feasible to collect, it may ultimately give less information than an optimized design. 

\subsection{Summary}

The paper is organized as follows. Section~\ref{sec:ch4background} is an overview of the relevant background material in sampling design, spatial statistics, and optimization. We review other optimal sampling methods; the spatial statistical methods that we rely on in our simulations; and optimization methods that can be used to solve our model. In Section~\ref{sec:ch4methods} we construct our model and outline how it differs from the existing literature. We show how our design criterion can be calculated as a MILP objective function, and discuss the design-based benchmarks that we use to test the efficacy of our model. We also discuss the details of the three increasingly complex logistics scenarios we use in the benchmarking study. In Section~\ref{sec:ch4results}, we present the results of our benchmarking study, and show example designs for each of the three scenarios, under each of the simulated design methods. Finally in Section~\ref{sec:ch4conclusions} we draw conclusions about the performance of our method and outline next steps.  

\section{Background}\label{sec:ch4background}
Spatial sampling design is a well-studied problem with a variety of existing methods. To place our method in context, we discuss: (1)  the set of design-based sampling alternatives; (2) the set of model-based optimal sampling alternatives; and (3) MILP models, their capabilities, and common methods of solving them. We then discuss in Section~\ref{sec:ch4methods} our specific model choices.

\subsection{Random Sampling Methods}

Simple random sampling (SRS) is the foundation of most random sampling methods. It consists of selecting samples independently at random from the set of all possible samples with equal probability of inclusion although unequal probabilities of inclusion is a natural extension \citep{lohr2019sampling}. However, one of the core principles of spatial statistics is the presence of autocorrelation \citep{tobler1970computer}. Spatial surveys are therefore often designed to account for this autocorrelation by ensuring that points are spatially balanced. Design-based inferences from SRS are still valid when locations are autocorrelated, but are generally less precise than if autocorrelation was accounted for. 

An alternative is stratified random sampling (StratRS). In this method, the space is divided into a set of disjoint subspaces (called strata), and each of these is sampled separately. This forces spatial balance on the large scale by ensuring points are evenly distributed between regions, but does not enforce spatial balance on the small scale, because balance within strata is not necessarily controlled. A third method, systematic sampling, similarly enforces large scale spatial balance, by sampling on a regular grid, often square, hexagonal, or triangular \citep{white1992cartographic} and often with a randomized starting point \citep{bellhouse2005systematic}. 

A more complex method of achieving spatial balance in a randomized method is through any of a number of so-called spatially balanced random sampling (SBRS) methods. SBRS methods include generalized random tesselation stratified (GRTS) sampling \citep{stevens2004spatially} and balanced acceptance sampling (BAS; \cite{robertson2013bas}). GRTS sampling operates by mapping two-dimensional space onto one-dimensional space, selecting a systematic sample of 1-D space with a random starting point, then mapping the sample back to two-dimensional space, while BAS leverages the Halton sequence to balance location in any number of dimensions. Both GRTS and BAS generally account for spatial autocorrelation equally well \citep{kermorvant2019optimizing}.

None of these methods take account of logistical or budgetary constraints, although all except systematic sampling are able to accommodate unequal inclusion probabilities, which can be used to adjust for (but not guarantee satisfaction of) a knapsack constraint (a single constraint assigning a cost to each sample and an overall budget). In the presence of more complex constraints feasibility cannot necessarily be guaranteed, and even if it can, the main way feasibility can be attained is by decreasing sample size and therefore quality of inference.

\subsection{Optimal Sampling Methods}

An optimization problem in general consists of two components: an objective function that one seeks to either minimize or maximize; and a set of constraints that must be met by the optimum. Although a problem can be defined with only these two components, a third component is also essential in practice: namely the choice of method by which the optimal solution will be found or approximated. 

Because optimization problems frequently include a set of non-trivial constraints, it is natural to consider using an optimization method when feasibility is a primary concern. However, in the case of optimal sampling design, most existing methods treat constraints as only a secondary consideration. 

Previous studies have focused on the choice of objective function (frequently termed ``design criterion'' in this context). Some common criteria are the $A$-, $D$-, $E$-, and $T$-criteria (Ch. 6, \cite{pukelsheim2006optimal}). These objective functions can be interpreted as minimizing, for example, the average variance of a set of model parameters (the $A$-criterion), or the determinant of a dispersion matrix (the $D$-criterion). They can also be interpreted more abstractly as ways of mapping matrices to ``sizes.'' The $D$-criterion can for example be defined in terms of an $n\times n$ matrix $\bm{M}$ as $\det(\bm{M})^{1/n}$, while the $A$-criterion can be defined as $(\frac{1}{n}\mbox{trace } \bm{M}^{-1} )^{-1}$. Each of these criteria is distinct and will in general imply a different optimal sample.

Our method relies on a variant of the $C$-criterion, which can be interpreted as minimizing the conditional variance of an average of model parameters, and can be expressed as an inner product $\bm{v}^T\bm{M}^{-1}\bm{v}$ for a constant vector $\bm{v}$. 
One challenge when applying the $C$-criterion is how to choose the vector $\bm{v}$. We propose that when a spatial function is represented by a vector of model parameters, such as in the SPDE approach, $\bm{v}$ can be chosen naturally to minimize the variance of an estimate of an areal mean. 

The set of constraints that can be applied depends upon the method being used to find or approximate a solution. For example, the popular Broyden-Fletcher-Goldfarb-Shanno algorithm recquires modification to be able to incorporate even simple box-type constraints \citep{byrd1995limited}. In general, integer constraints are particularly challenging to apply, because the domain becomes discontinuous. In the context of optimal sampling, a variety of methods have been applied. For example, in a small design space (such as when only a single location is to be sampled) the optimal sample may be found by enumeration (e.g. \cite{leach2022recursive}). Such a method can be extended to produce a greedy algorithm which sequentially selects the best single location to add to a sample (e.g. \cite{wynn1970sequential,diaz2018sparse}). In certain cases, the greedy algorithm may even have provably near-optimal performance, although in these cases the set of modelable constraints is often limited, for example to knapsack and matroid-type constraints \citep{krause2008near,lee2009non}. Other models also admit knapsack-type constraints, such as the Elfving method \citep{elfving1952optimum,stoica2010algebraic}. For complex models, methods such as genetic algorithms (e.g. \cite{dupont2021optimal}) or simulated annealing (e.g. \cite{van1999constrained}) can be used to produce high-quality designs, under logistical constraints, although these generally provide no guarantee as to the quality of the design relative to the optimum.  


\subsection{Mixed Integer Linear Programming}

\subsubsection{Problem Formulation}
Linear programming encompasses the methods used to solve optimization problems of the following form, termed linear programs (LPs):
\begin{mini}
{\bm{x}}{\bm{c}^T\bm{x},}
{\label{eq:LP-example}}{}
\addConstraint{\bm{A}^T\bm{x}}{\leq \bm{b}.}
\end{mini}
Many common optimization problems can be cast as LPs, including the knapsack problem, the diet problem, the multicommodity flow problem, and problems involving manufacturing of multiple types of goods (Chapter 1; \cite{bertsimas1997introduction}). Linear programs are therefore used frequently in industrial applications, where they are solved using industrial solvers such as CPLEX \citep{cplex2020user} and Gurobi \citep{gurobi}. In practice, they are often solved by the simplex method.

Logistics involving integer or binary constraints in addition to linear constraints cannot be modeled using linear programs. These problems may instead be modeled using mixed integer linear programs (MILPs). The addition of integer and binary constraints allows a wider range of logistics to be modeled including among many others the traveling salesman problem, and a variety of more complex manufacturing, scheduling, and transportation problems. Our model is a MILP and can therefore accommodate these and many other logistical constraints.

\subsubsection{Branch and Bound}
MILPs are difficult to solve computationally (being in general NP-hard), but due to their importance in industry, techniques have been developed to find high-quality solutions nevertheless. By default industrial solvers such as CPLEX and Gurobi use the branch and bound algorithm \citep{land2010automatic}, which allow these solvers to both find high-quality solutions, and calculate bounds on how much better the true optimal solution could be. 

Branch and bound (B\&B) is the main algorithm used in the solution of NP-hard combinatorial optimization problems \citep{clausen1999branch}. In general, B\&B can be seen as more of a paradigm than a specific algorithm, however in the case of MILPs, the industrial solvers have developed tunable B\&B algorithms that work well on a wide variety of problems. The B\&B paradigm generally works by dynamically generating a tree whose root is the set of all possible solutions and whose branches are subsets thereof. Each time a branch is generated (often by fixing the value of a particular discrete variable), a function is evaluated to bound the objective function values that may be obtained within that branch. In the case of MILPs, the bounding function is usually a relaxation of the integrality constraints to produce a more easily solved LP, the solution of which provides a bound on the MILP solution. Periodically, heuristic methods are used to generate solutions from within the various branches as they are explored, and the best solution is maintained at any given time. Often, a so-called ``warmstart'' solution which is expected to be of reasonably high-quality is provided at the start of the algorithm. Solutions are compared to bounds, and when a solution is found of higher quality than a branch's bound, the branch is pruned from the tree. Over time it is often possible to prune large portions of the tree, and to dynamically prove a global bound on the optimal solution \citep{land2010automatic}. Comparing this global bound with the current candidate solution provides a dynamically updated ``optimality gap'' often expressed in both absolute and relative terms. This ability to improve the optimality gap over time, until either a certain gap is achieved or until a time limit is reached, is an appealing property of the B\&B paradigm.

\subsubsection{Linearizing a Binary-Continuous Product}
Our formulation involves calculating the product $z_i$ of a binary decision variable $x_i$ with a linear combination of continuous variables $\bm{a}_i^T\bm{y}$. This operation, although non-linear, can be carried out within a MILP through the use of additional constraints (Chapter 9; \cite{williams2013model}), so long as the range of $\bm{a}_i^T\bm{y}$ is bounded both above and below. This is accomplished by prescribing the following set of standard constraints:
\begin{align*}
    z_i & \leq \bm{a}_i^T\bm{y} - (1-x_i)L, \\
    z_i & \geq \bm{a}_i^T\bm{y} - (1-x_i)U, \\
    z_i & \leq x_iU, \\
    z_i & \geq x_iL.
\end{align*}
When $x_i$ is equal to 1 the latter two constraints simply enforce the \emph{a priori} upper and lower bounds $L \leq \bm{a}_i^T\bm{y} \leq U$, while the former two enforce the desired equality $z_i = \bm{a}_i^T\bm{y}$. Meanwhile, when $x_i$ is equal to 0, the latter two enforce the desired equality $z_i = 0$ while the former two bounds are loose because again, by the \emph{a priori} bounds $0\leq \bm{a}_i^T\bm{y} - L$ and $0\geq \bm{a}_i^T\bm{y} - U$. This technique further expands the set of possible problems which may be modeled using MILPs to include those with bounded binary-linear products.

\section{Methods}\label{sec:ch4methods}

\subsection{Design Criterion}

Consider a set of parameters $\bm{u}$, jointly modeled as multivariate normal with precision matrix $\bm{Q}$. We assume without loss of generality that the prior mean is 0. Suppose that a set of linear combinations $\bm{a}_i^T\bm{u}$ are candidates to be sampled, and that the set of observations $d_i = \bm{a}_i^T\bm{u} + \epsilon_i$ are obscured by a vector of i.i.d Gaussian noise $\bm{\epsilon}$, with known variances $\sigma^2_i$. Finally, we assume that the object of inference is a known linear combination $\bm{v}^T\bm{u}$. Our design criterion is the conditional variance $C^2$ of $\bm{v}^T\bm{u}$ given a the chosen set of observations $\bm{d}$.

Under this formulation, our model can be written as follows, where $x_i$ is a set of indicator variables denoting whether the observation $d_i$ is sampled:
\begin{align}
\bm{d} & = \bm{A}^T\bm{u} + \bm{\epsilon} \\
\bm{u}|\bm{d} & = N\left(0,\bm{Q} + \sum_{i=1}^n \frac{x_i}{\sigma^2_i}\bm{a}_i\bm{a}_i^T\right) \\
C^2 & = \bm{v}^T\left(\bm{Q} + \sum_{i=1}^n \frac{x_i}{\sigma^2_i}\bm{a}_i\bm{a}_i^T\right)^{-1}\bm{v}.
\end{align}

In the case of spatial sampling, the vector $\bm{u}$ may represent a Gaussian random field $u(\bm{s})$ over a spatial coordinate $\bm{s}$, discretized according to the SPDE method \citep{lindgren2011explicit}, in which case the vector $\bm{v}$ may be chosen to approximate a spatial integral over $u(\bm{s})$, and the precision matrix $\bm{Q}$ is sparse. Modifications to our method to make use of covariance matrices and to allow uncertainty in the covariance function are discussed in Appendix~3. 

Our objective function is a Bayesian formulation of the $C$-criterion, which would more typically use the Moore-Penrose pseudoinverse of an information matrix (e.g. \cite{sagnol2010optimal}) rather than a precision matrix. Using our formulation, we may linearize our objective function by defining the following set of intermediate variables:
\begin{align}
\bm{y} & = \left(\bm{Q} + \sum_{i=1}^n \frac{x_i}{\sigma^2_i}\bm{a}_i\bm{a}_i^T\right)^{-1}\bm{v} \\
z_i & = x_i(\bm{a}_i^T\bm{y}),
\end{align}
so that our objective function may be written as $\bm{v}^T\bm{y}$, under the constraint
\begin{equation}
\bm{Q}\bm{y} + \frac{1}{\sigma^2}\bm{A}^T\bm{z} = \bm{v} .
\end{equation}
This allows us to optimize the decision variables $x_i$ using a MILP.

\subsection{Bounds}

As outlined in Section~\ref{sec:ch4background}, to calculate the product variables $\bm{z}$ it is necessary for all elements of $\bm{A}^T\bm{y}$ to be bounded above and below. Such bounds may be derived using the Cauchy-Schwartz inequality, because the precision matrix $\bm{Q} + \sum \frac{x_i}{\sigma_i^2}\bm{a}_i\bm{a}_i^T$ defines an inner product $\inner{\bm{u},\bm{v}}_{X} = \bm{u}(\bm{Q}+\sum \frac{x_i}{\sigma^2_i}\bm{a}_i\bm{a}_i^T)^{-1}\bm{v}$ such that $\inner{\bm{u},\bm{u}}_{\bm{x} + \bm{\zeta}} \leq \inner{\bm{u},\bm{u}}_{\bm{x}}$, where $\bm{\zeta}$ and $\bm{x} + \bm{\zeta}$ are both 0/1-vectors. Thus, 
\[
z_i^2 = \inner{\bm{a}_i,\bm{v}}_{\bm{x}}^2 \leq \inner{\bm{a}_i,\bm{a}_i}_{\bm{x}}\inner{\bm{v},\bm{v}}_{\bm{x}} \leq \inner{\bm{a}_i,\bm{a}_i}_{0}\inner{\bm{v},\bm{v}}_{0}.
\]
In fact we may strengthen these bounds by noting that $\inner{\bm{v},\bm{v}}$ is our objective function, so that at optimality 
\[
z_i^2 \leq \inner{\bm{a}_i,\bm{a}_i}_{\bm{x}_{\rm opt}}\inner{\bm{v},\bm{v}}_{\bm{x}_{\rm opt}} \leq \inner{\bm{a}_i,\bm{a}_i}_{0}\inner{\bm{v},\bm{v}}_{\bm{x}_{\rm feas}}.
\]
Thus the bounds on $z_i$ can be strengthened whenever an improved feasible solution is found. To allow a single model to be run, we use our greedy algorithm solution for $\bm{x}_{\rm feas}$ and do not update this further. 

We strengthen the bounds further by using two different sets of bounds in the linearization equations
\begin{align*}
L^{(0)}_ix_i & \leq z_i \leq U^{(0)}_ix_i, \\
\bm{a}_i^T\bm{y} - U^{(1)}_i(1-x_i) & \leq z_i \leq \bm{a}_i^T\bm{y} - L^{(1)}_i(1-x_i),
\end{align*}
namely, an upper and lower bound when $x_i = 0$ and an upper and lower bound when $x_i = 1$. Specifically, when $x_i = 1$ we may strengthen our bounds using the Sherman-Morrison formula
\[
\inner{\bm{a}_i,\bm{a}_i}_{\bm{e}_i} = \inner{\bm{a}_i,\bm{a}_i}_{0} - \frac{\inner{\bm{a}_i,\bm{a}_i}_{0}^2}{1 + \inner{\bm{a}_i,\bm{a}_i}_{0}} = \frac{\inner{\bm{a}_i,\bm{a}_i}_{0}}{1 + \inner{\bm{a}_i,\bm{a}_i}_{0}}.
\]
where $\bm{e}_i$ is an index vector. Therefore our bounds $L^{(1)}_i$ and $B^{(1)}_i$ can be strengthened by a factor of $1 + \inner{\bm{a}_i,\bm{a}_i}_0$ to
\[
z_i^2 \leq \inner{\bm{a}_i,\bm{a}_i}_{\bm{e}_i}\inner{\bm{g},\bm{g}}_{\rm feas} = \frac{\inner{\bm{a}_i,\bm{a}_i}_{0}\inner{\bm{g},\bm{g}}_{\rm feas}}{1+\inner{\bm{a}_i,\bm{a}_i}_{0}}.
\]
Empirically these bounds appear to be relatively weak, and strengthening them further is a promising area of future study.



\subsection{Benchmarking}

\subsubsection{Logistics Scenarios}
For the purposes of benchmarking, we formulated a series of three increasingly complex logistics models, which we call Knapsack, Helipad, and Tanana. Our Knapsack model assigns to each possible observation a fixed cost $c_i$, and imposes the constraint $\bm{c}^T\bm{x} \leq B$, for budget $B$. In our demo, these costs are generated based on two-dimensional location $(s,r)$ from the linear function $c(r,s) = r+s+1$. The budget is set to be 100. Both the Knapsack and Helipad scenarios were simulated on the unit square. 

Our Helipad model maintains the knapsack constraint, but additionally assumes that observations must be accessed from a series of 9 equally spaced helipads, each with a limited range. Each location is therefore only accessible from a small number of nearby helipads, and to access a point from a given helipad, the helipad must be maintained for the season, incurring a fixed cost. It is therefore possible to save costs by using fewer helipads overall, thereby forgoing the locations that are accessible from those locations, but allowing more locations to be sampled from the helipads which are maintained. 

The helipad constraint can be modeled by an adjacency matrix $\bm{D}$ describing whether point $i$ is reachable from helipad $j$. Then the following set of constraints can be used, using the vector $\bm{h}$ for the helipad binary decision variables, and the vector $\bm{f}$ for the fixed costs of maintenance:
\begin{align*}
\bm{c}^T\bm{x} + \bm{f}^T\bm{h} \leq B \\
x_i \leq \sum \bm{D}_{ij}\bm{h}_j. \\
\end{align*}

In our demonstration, the cost of each helipad is set to be a constant $f = 10$ for all helipads. The cost of sampling was again set to be $c(r,s) = r+s+1$. The range of each helipad was set to be $\frac{1}{3\sqrt{2}}$ to allow only a small amount of overlap between ranges on the unit square. 

Finally, our Tanana model was developed in collaboration with the logistics team for the US Forest Service's (USFS) Forest Inventory Analysis (FIA) project in Alaska; in which a major goal is the estimation of total carbon sequestration. This model matched actual on-the-ground logistics as closely as possible, using the actual geometry of the region. Our Tanana logistics model included: the actual set of available helicopter bases, fuel costs, costs of food and lodging, variable efficiency in number of plots that can be sampled per day from each base, a fixed maximum number of days in the sampling season of each of multiple years, and a fixed cost of lost sampling days incurred by traveling to a new base mid-season. The complete set of logistics modeled are described in Table~\ref{tab:logistics}. 

The Tanana logistics motivate the following constraints:
\begin{align*}
O_i + F_i & \leq B_i && \mbox{operating cost + flight cost $\leq$ budget}\\
H_i + S_i & \leq D_i && \mbox{hub relocation + sampling $\leq$ season length}\\
O_i & = \sum_j c_{j}s_{i,j} && \mbox{operating cost = $\sum$ cost of site$\times$days at site}\\
F_I & = \sum_j f_jd_{i,j} && \mbox{flight costs = $\sum$ fuel cost at site$\times$distance traveled}\\
H_i & = 2*\left(\sum_j h_{i,j} - 1\right) &&\mbox{hub relocation days = 2 per hub, except the first}\\
S_i & = \sum_{j} s_{i,j} &&\mbox{sampling days = $\sum$ sampling days per site} \\
p_{i,j} & = \sum_{k}x_{i,j,k} && \mbox{plots sampled at site = $\sum$ plot indicator variables} \\
s_{i,j} & = p_{i,j}/e_j && \mbox{sampling days = plots sampled / plots per day}\\
d_{i,j} & = \sum_k 4 d'_{j,k}x_{i,j,k} && \mbox{distance traveled = $\sum$ 4$\times$distance to sites sampled}\\
h_{i,j} & \leq M_{k,j}x_{i,j,k} \forall i,j && \mbox{a hub must be used if a site is sampled from it}\\
x_{i,j,k} & \leq M_{k,j}h_{i,j} \forall i,j,k && \mbox{if a site is sampled, it must be from some hub}
\end{align*}
where $B_i$ and $D_i$ describe the budget and days in season $i$, which are generally constant but are reduced by the time and money spent conducting the pilot study in the first year; $O_i$ represents the operating cost in year $i$; $F_i$ represents the annual flight costs; $H_i$ represents the number of days spent transferring hubs in year $i$; $S_i$ describes the number of days spent sampling in year $i$; $c_j$ describes the daily cost of operations at hub $j$; $s_{i,j}$ denotes the number of days spent sampling at hub $j$ in year $i$; $f_j$ denotes the fuel costs at hub $j$; $d_{i,j}$ denotes the distance flown from hub $j$ in year $i$; $h_{i,j}$ denotes whether hub $j$ was used in year $i$; $x_{i,j,k}$ denotes whether plot $k$ was sampled from hub $j$ in year $i$; $p_{i,j}$ denotes the number of plots sampled from hub $j$ in year $i$; $e_j$ denotes the efficiency of plot $j$ measured in plots/day; and $d'_{j,k}$ denotes the distance from hub $j$ to plot $k$. To facilitate subsampling of the actual set of actual sampled locations, we used only two sampling seasons, each of reduced length and budget. 

\begin{table}[htp]
    \centering
    \begin{tabularx}{\textwidth} { 
   >{\raggedleft\arraybackslash\hsize=.4\hsize}X 
   >{\raggedright\arraybackslash\hsize=1.6\hsize}X  }
 Annual budget & Each season must be individually within budget. We used $1/5$th the true budget to accommodate sub-sampling the actual locations. \\ \hline
Length of sampling season & Each season must be completed in a fixed number of days. We used $1/5$th the true season to accommodate sub-sampling the actual locations. \\ \hline
Helicopter hubs & Teams of samplers lodge near hubs, and are flown by helicopter to nearby sites within range. \\ \hline
Multi-team routing & Multiple teams share a single helicopter, which must drop off a single team before returning to transport the next. The helicopter then must pick up each team at the end of the day. \\ \hline
Variable efficiencies & Due to a wide variety of factors, hubs have a variable average number of sites visited per day \\ \hline
Variable cost of operation & Due to a wide variety of factors, hubs have a variable daily cost of operation \\ \hline
Cost of fuel & Each mile traveled incurs a $4\times$ fuel cost to drop off the team, return, pick them up, and return. \\ \hline
Hub switching & Every time the teams move to a new hub they lose 2 days of productivity \\ 
\end{tabularx}
\caption{Logistic factors modeled in the Tanana scenario.}\label{tab:logistics}
\end{table}

\subsubsection{Quantifying Efficacy}

To estimate the efficacy of our method, we simulated the full process by which such a method would be implemented in practice. First we generated the underlying spatial field $\bm{u}$ as a Mat\'{e}rn Gaussian process with parameters as outlined in Table~\ref{tab:parameters}. We chose the $\alpha$ parameter to ensure our simulations reflected the reality that $\alpha$ is frequently unidentifiable (i.e. because our models assumed $\alpha = 2$ instead of 4; \cite{lindgren2011explicit}). We set $\sigma^2_\epsilon$ to one for all scenarios. We converted between $\kappa$ and $\rho$ using the empirical relationship $\kappa = \sqrt{8\nu}/\rho$ \citep{lindgren2011explicit} and chose both so that the range was smaller than but on the same order of magnitude as the spatial extent of the domain. In this regime, spatially balancing can be expected to perform reasonably well, allowing us to treat our results as a sort of baseline level of performance for our method relative to SBRS. If range was either very large or very small, it would be advantageous to cluster points either in the cheapest locations, or nearby to the edges, which would tend to widen the gap between our method and the benchmarks. We chose $\sigma^2$ to reflect both a high-uncertainty regime for the Knapsack and Helipad scenarios and a low-uncertainty regime for the Tanana scenario, in which great expense is taken to reach a location and therefore high quality data are taken at each location.

\begin{table}[]
    \centering
    \begin{tabular}{c cccccc}
        Model & $\alpha$ & $\sigma^2$ & $\kappa$ & $\rho$ & $\sigma^2_\epsilon$ & Time Limit  \\ \hline
        Knapsack & 4 & 20 & 2 & $\approx 0.24$ & 1 & 5min \\
        Helipad & 4 & 20 & 2 & $\approx 0.24$ & 1 & 20min \\
        Tanana & 4 & 1000 & $\approx 1.63\times 10^{-5}$ & 300 km & 1 & 4hr  
    \end{tabular}
    \caption{The parameter values used for the three different scenarios of our benchmarking study. Greek letters refer to parameters used to simulate the data-generating process. We translated between $\kappa$ and $\rho$ using the empirical equation from \citep{lindgren2011explicit}. Time Limit refers to the time limit at which the Gurobi solver terminates the B\&B procedure and returns the best known solution.}
    \label{tab:parameters}
\end{table}

To model the requisite pilot study to estimate the covariance function hyperparameters, we used sampling routines designed to balance estimation of these hyperparameters with budget and feasibility considerations for the pilot study, and with optimality of the eventual sample for estimating the areal mean. For the Helipad and Tanana scenarios, we chose two bases with which to collect the pilot study. In the Tanana scenario, due to wide variation in the area covered by each base, these bases were fixed to be the two with the largest covered area. In the Knapsack and Helipad scenarios, pilot samples were then chosen according to an incremental routine designed to produce some clusters and some distant points, by weighting points according to their distance from existing points, then sampling randomly. In the Tanana scenario, due to the complexity of the logistics, the need to overestimate the range was more pressing, and we therefore conducted a SBRS procedure to collect the pilot study, because precise estimation of the hyperparameters was less of a priority. 

In all cases, the selected pilot sample was then used to produce simulated data, by adding i.i.d Gaussian noise to the simulated spatial field. We used these data to fit the hyperparameters using maximum \emph{a posteriori} estimation, assuming for the sake of testing robustness an \emph{incorrect} value for the hyperparameter $\alpha$, which is difficult to identify in practice, and is frequently fixed \emph{a priori} \citep{lindgren2011explicit}. It was necessary to obtain a point estimate in order to have a single matrix $\bm{Q}$ to use in the final optimization model. An important hyperparameter for our model is the range (determined by $\kappa$ in the Mat\'{e}rn model), because if the range is small the model attempts to save costs by moving observations closer together, which it can do while sacrificing only very little information, because when the range is small, observations are approximately uncorrelated even at low distances. When the range is underestimated, however, this can be detrimental to the resulting inference, thus to produce a conservative design we imposed informative priors overestimating the range. 

After simulating a pilot study, we generate a warmstart sample using a greedy algorithm. A warmstart is an initial guess at a near-optimal solution, often one which is known to be feasible, and provides the B\&B algorithm a heuristic from which to begin solving. In our case we use a greedy algorithm, in which we evaluate the decrease in our objective function for each candidate location then sort the candidates by the size of this decrease, and evaluate feasibility until we find the most informative feasible location. Feasibility is determined using an auxiliary MILP model, generated by removing the objective function and the variables $\bm{y}$ and $\bm{z}$ used to calculate it. When no locations are feasible the algorithm terminates. This warmstart was provided to Gurobi as a heuristic solution, then Gurobi was allowed to run for a fixed period of time, which was longer for more complex scenarios. At the end of this period of time, the algorithm outputs the best sample found. Due to the size of our simulation, the B\&B algorithm was necessarily constrained by a relatively short time limit. If a more optimal solution was able to be found after a longer period of time, we expect the gap between our method and the benchmark comparisons would improve, as would the reported optimality gap.

For comparison, we also conducted simple random sampling, stratified random sampling (except for Tanana, which featured a less regular domain), and spatially-balanced random sampling (SBRS). In the Knapsack and Helipad models, these methods were allowed to choose points continuously from the domain, and SBRS was accomplished using balanced acceptance sampling (BAS; \cite{robertson2013bas}). In these scenarios, candidate locations for optimization were generated as a dense square grid. In the Tanana scenario, the set of possible observations for the design-based methods were chosen from the actual, hexagonally systematic sample used by the US Forest Service (USFS) to sample the region, and we used Generalized Random Tesselation Stratified (GRTS) sampling for our SBRS method to accommodate the discrete set of candidate points \citep{stevens2004spatially}. For each design-based method, we used a series of sample sizes which ranged from near-certain infeasibility to near-certain feasibility. We also evaluated the performance of our greedy algorithm warmstart samples, as well as our B\&B samples, optimized using the Gurobi solver \citep{gurobi}.

For each method we then conducted the appropriate respective calculations to infer the areal mean of the simulated function. In the case of the greedy and optimized designs we used kriging to estimate the areal mean; we did not update the fitted autocorrelation model, but when not analyzing large numbers of simulations, a less programmatic analysis would be appropriate. We did not model covariates in any of our simulations to have a cleaner comparison to the base versions of our design based benchmarks, but show how one could do so in the Appendix~1. In the case of the design based methods we used the mean of the measurements, because we used equal weights across the whole spatial domain, or in the case of stratified random sampling the mean of the strata estimates (which were themselves of equal area).

Finally, we calculated the squared error for each simulation, as well as the feasibility. We bootstrap resampled these two benchmarks 1000 times to produce confidence intervals of mean squared error (MSE) and probability of feasibility. Our total number of simulations for each benchmark and scenario can be found in Table~\ref{tab:numsims}. For purposes of computation, we used fewer simulations for the more complex logistics models, because these took substantially longer to run. We allowed the Knapsack optimization routine to run for 5 minutes per simulation; the Helipad routine to run for 20 minutes per simulation; and the Tanana routine to run for 4 hours per simulation. 

\begin{table}[]
    \centering
    \begin{tabular}{c ccc}
          & Knapsack & Helipad & Tanana  \\ \hline
  Design  & 3000     & 1500    & 1000    \\
Optimized & 300      & 100     & 50      
    \end{tabular}
    \caption{The number of benchmarking simulations run to estimate MSE and probability of feasibility for each method, by scenario.}
    \label{tab:numsims}
\end{table}

\section{Results}\label{sec:ch4results}

\begin{figure}
    \centering
    \includegraphics[width=0.75\textwidth]{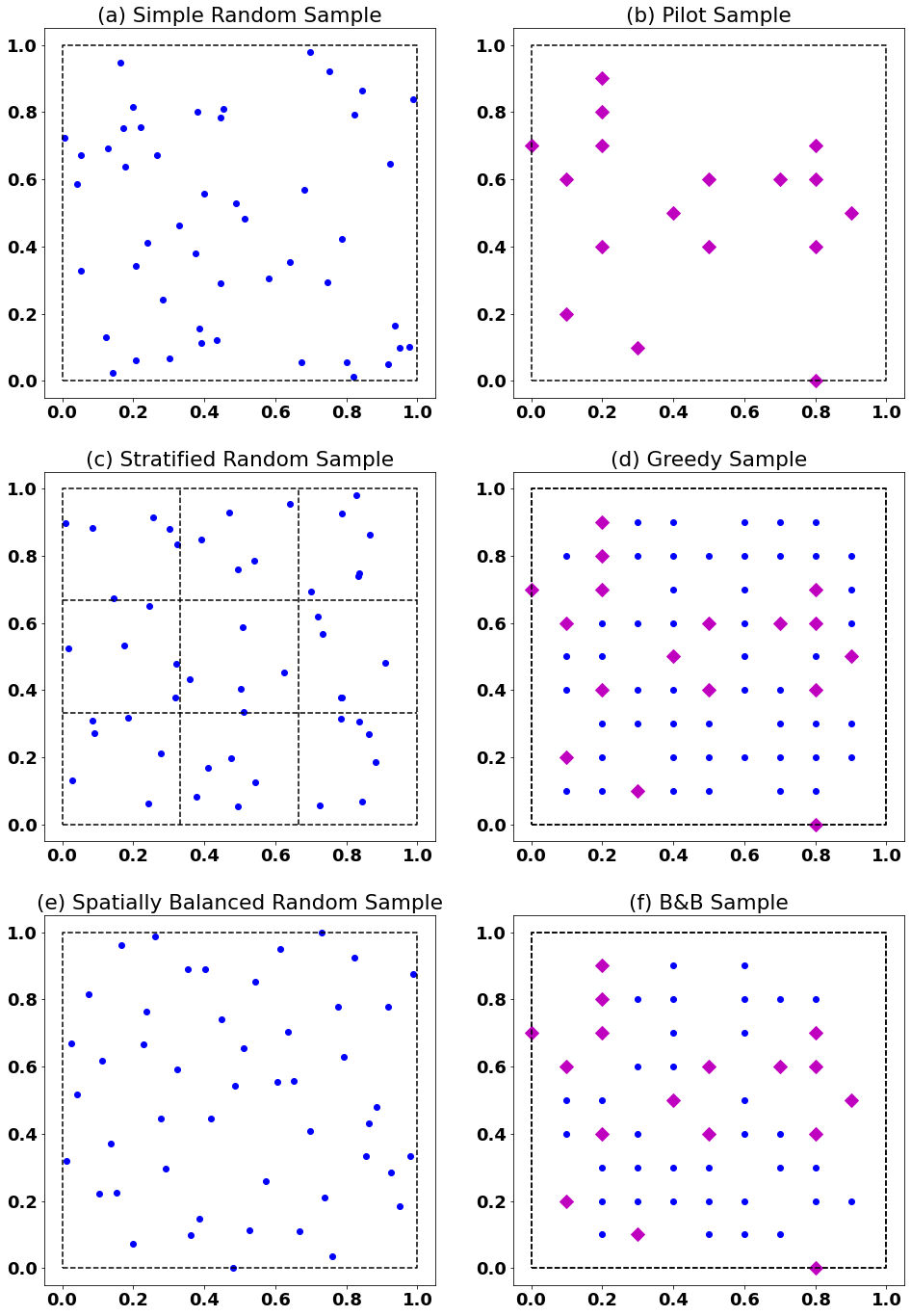}
    \caption{Example sample for the Knapsack scenario. Sample locations shown in circles (blue). Pilot samples shown in diamonds (magenta). Domain and strata outlined in dashed line (black). }
    \label{fig:knapsack-ex}
\end{figure}

\begin{figure}
    \centering
    \includegraphics[width=0.75\textwidth]{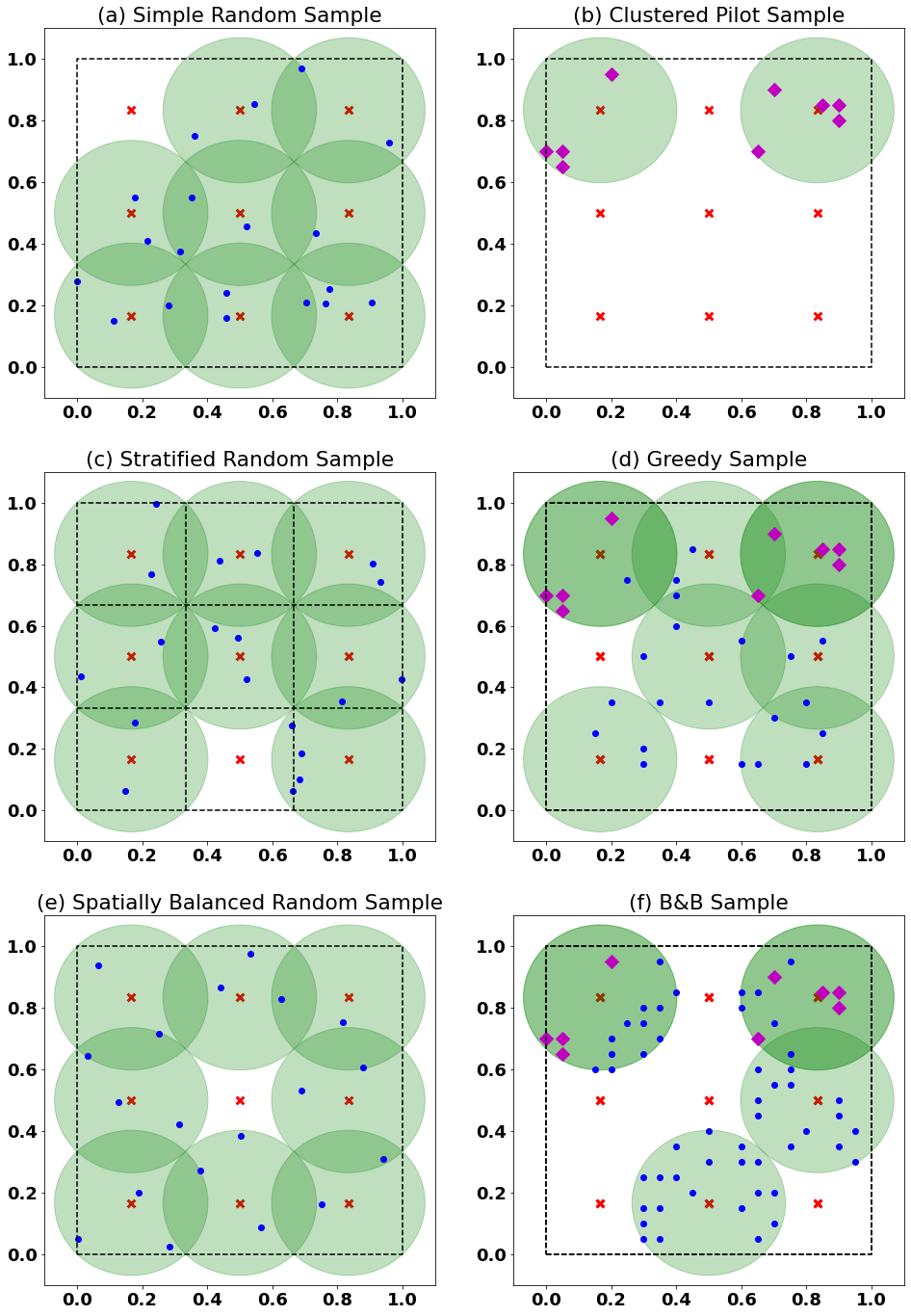}  
    \caption{Example sample for the Helipad scenario. Sample locations shown in circles (blue). Pilot samples shown in diamonds (magenta). Helipad locations shown in Xs (red). Helipad ranges shown in large transparent circles (green). Domain and strata outlined in dashed line (black). Choice of helipads for (a)-(e) determined by cost-minimization MILP.
    }
    \label{fig:helipad-ex}
\end{figure}

\begin{figure}
    \centering
    \includegraphics[width=0.75\textwidth]{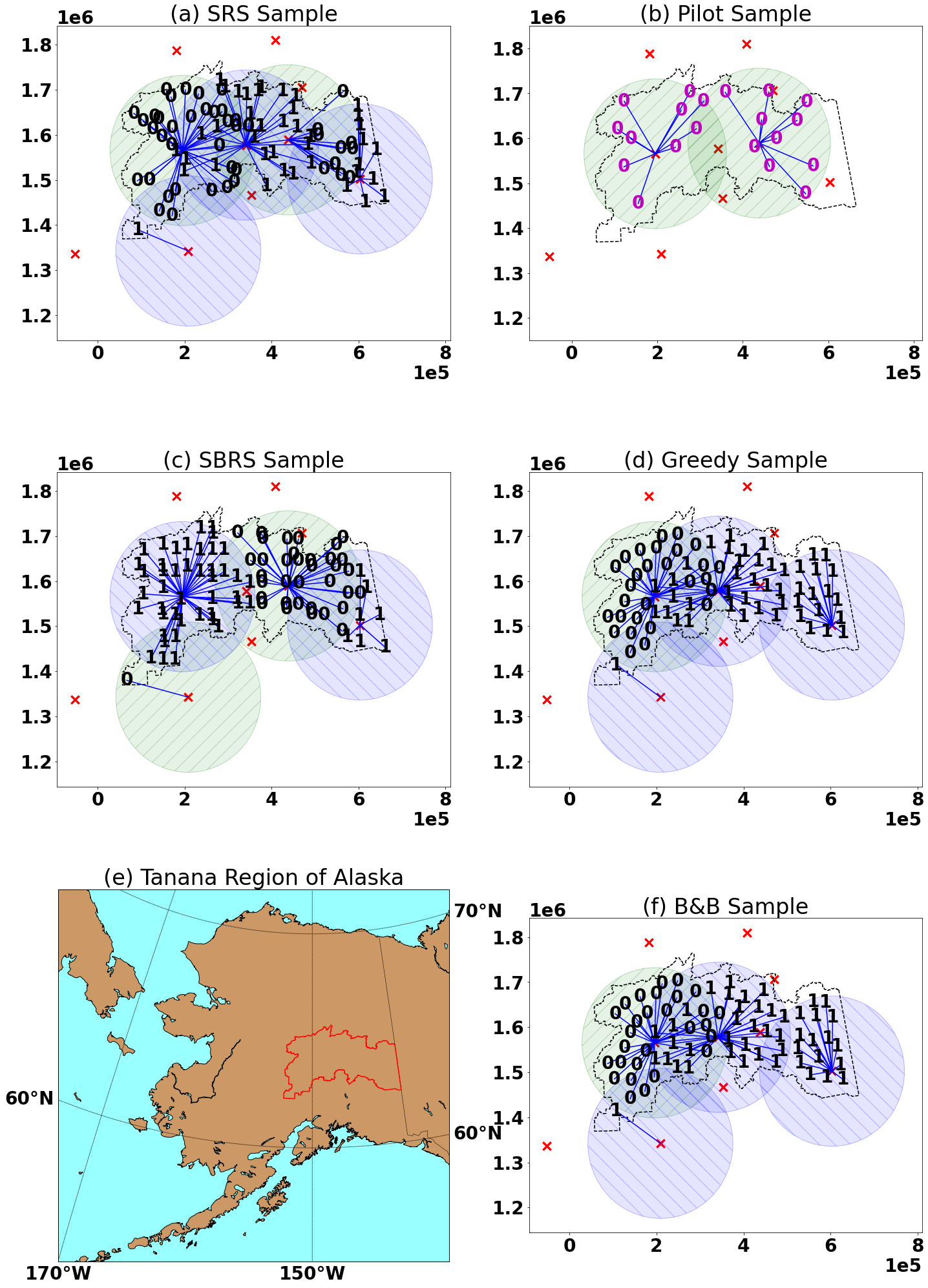}  
    \caption{Example sample for the Tanana scenario. Sample locations marked with 0 or 1 depending on which sampling season the sample is to be collected. Base ranges shown in large transparent circles (green with hatching toward upper-right for season 0; blue with hatching toward upper-left for season 1). Base used to reach each sample shown in radiating lines (blue). Domain (the Tanana region of Alaska) outlined in dashed line (black), and also outlined (red) in a map of Alaska in panel (e). Axes in panels (a)-(d) and (f) are in NAD83 / Alaska Albers coordinates; panel (e) is in latitude and longitude.}
    \label{fig:tanana-ex}
\end{figure}

Example designs produced by each of our benchmark methods, our pilot methods, our greedy warmstart, and our B\&B solution can be found in Figures~\ref{fig:knapsack-ex}-\ref{fig:tanana-ex}. Visually, our methods appear to space points out comparably to stratified sampling or SBRS. However, by accounting for logistics, we hypothesized that our method would be able to select a higher sample size of points while maintaining feasibility. 

We found that, for the Knapsack scenario, our model performed comparably to stratified random sampling, but underperformed relative to SBRS, which we hypothesize to  be due to the non-optimality of the pilot study. Knapsack logistics are well-studied and generally simpler logistics than our method is intended for; other optimal design methods such as submodular optimization may be more appropriate in these scenarios because the simpler structure allows for better optimization properties, such as optimality approximation guarantees \citep{lee2009non} on greedy samples. Indeed for knapsack logistics, it is even possible to produce a set of design-based optimal weights, provided that spatial covariance is not modeled (see Appendix~4). 

\begin{figure}
	\vspace{-3cm}
    \centering
    \includegraphics[width=0.7\textwidth]{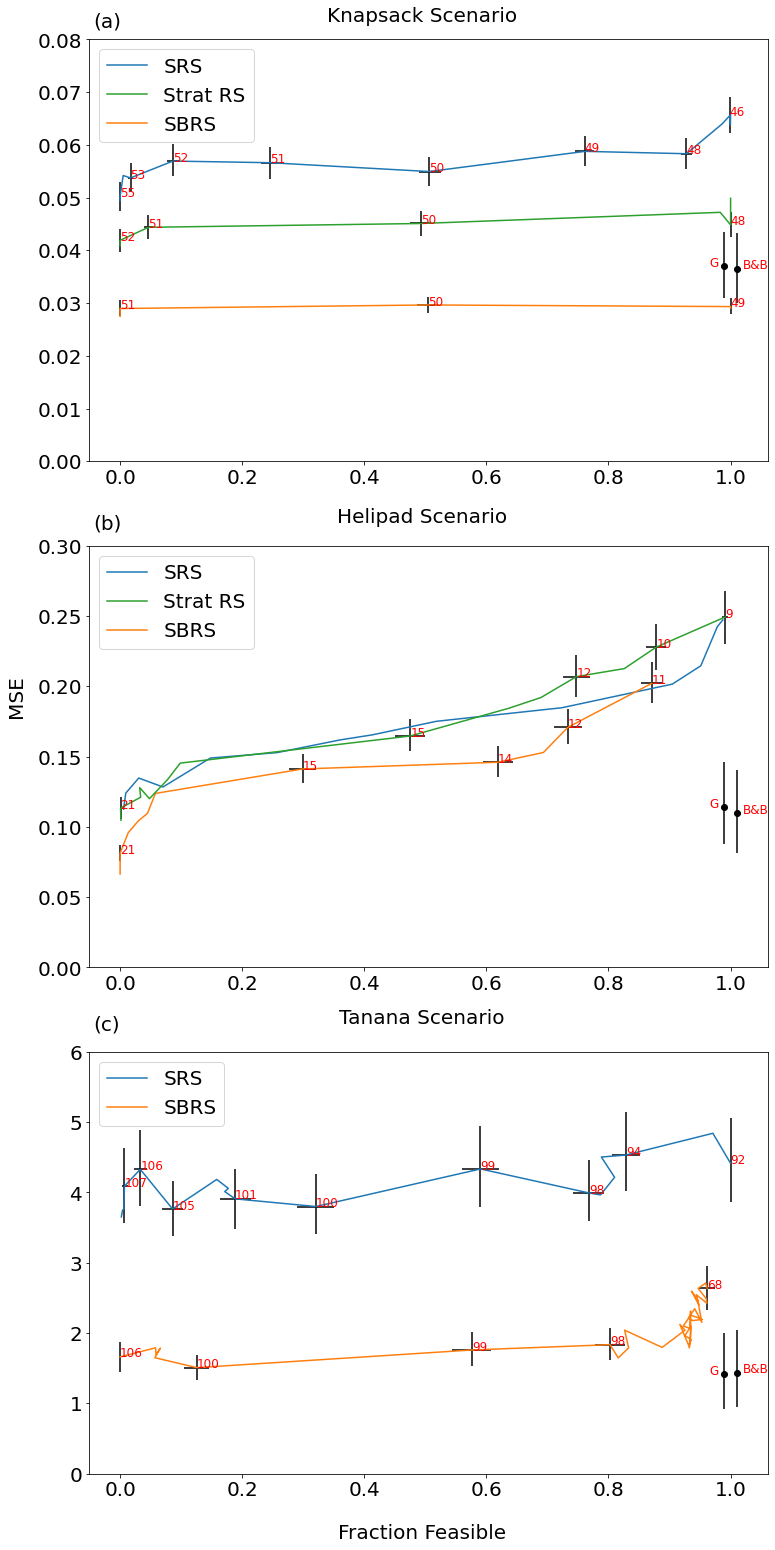}
    
    \caption{Results of our benchmarking simulations. Left-side plots contain the same lines as the right-side, but without the addition of bootstrapped error bars which obscure the underlying lines. Red numbers display the sample size for each point, and the characters ``B\&B'' and ``G'' stand for the branch and bound/greedy methods respectively.}  
    \label{fig:benchmarks}
\end{figure}

For more complex logistics, including both the Helipad and Tanana scenarios, both our greedy and optimized designs outperformed the design-based benchmarks. For the Tanana scenario, the logistics were sufficiently complex that even at small sample sizes it was difficult for the design-based methods to achieve near-certain feasibility. In contrast, our model produced a high-quality MSE comparable to relatively large sample-size SBRS, while guaranteeing feasibility. Total simulation results can be found in Figure~\ref{fig:benchmarks}, where we show how the MSE and probability of feasibility change with sample size for each of our competing methods, as well as bootstrapped confidence intervals showing the uncertainty in both metrics. The greedy and branch and bound solution metrics are displayed as points, because we cannot vary the sample size. 

Proportional improvements for the branch and bound solution relative to benchmarks are shown in Table~\ref{tab:improvements} as percentages of the benchmark MSE, calculated from bootstrapped confidence intervals. Note additionally that only in the Knapsack scenario did all benchmarks achieve 100\% simulated feasibility. Although it is more difficult to calculate precise estimates, it is illustrative also to consider the loss of feasibility that would be recquired to match the MSE of our model solutions. Comparing horizontally the extremes of our bootstrap intervals, for the Helipad scenario, recquisite losses appear to range from $\approx$30-95\% and for the Tanana scenario $\approx$5-100\% probability of feasibility would need to be sacrificed. 

\begin{table}
\centering
\begin{tabular}{c ccc}
        & Knapsack & Helipad & Tanana  \\ \hline
SRS     & 44-54\%  & 56-67\% & 68-79\% \\
StratRS & 19-32\%  & 56-67\% &  NA     \\
SBRS    & -4-25\%  & 45-59\% & 45-64\%
\end{tabular}
\caption{Proportional improvements in MSE demonstrated by our B\&B solution as a percentage of our three benchmarks' MSEs. For SBRS in the Knapsack scenario, shows the proportional increase in MSE. StratRS was not used in the Tanana scenario.}\label{tab:improvements}
\end{table}

We also observed relatively little difference between our B\&B solution and our greedy warmstart solution. The MILP optimization routine outperformed the greedy heuristic routine on average, but by a relatively small margin. Meanwhile, the B\&B routine required a substantially larger computational burden. 

\section{Conclusions}\label{sec:ch4conclusions}

For even moderately complex logistics scenarios, our two optimization routines outperformed all of our design-based benchmarks, despite making model choices that would disfavour our routines, including: assuming an incorrect value for the $\alpha$ parameter, frequently overestimating the range, and not updating the hyperparameters after sampling. These results are in line with a recent study comparing design-based and model-based methods for spatial sampling and inference more broadly, which found that even when model assumptions are violated, model-based approaches slightly outperformed SBRS and greatly outperform SRS \citep{dumelle2022comparison}. The use of a logistics-aware sampling design method has the potential to ease implementation of large-scale complex designs, by shifting complexity from the often time-sensitive implementation stage to the planning stage. We expect logistics-aware sampling design to be advantageous in remote regions such as Alaska, or in surveys involving marine or aerial support, where transport and routing may account for a substantial fraction of overall cost. Using a model-based optimal design approach can ensure that limited time and money are spent efficiently to obtain the highest amount of information possible. 

In the three scenarios we evaluated, the greedy heuristic performed comparably to the MILP optimization routine, while requiring substantially lower computation burden. In this particular set of scenarios therefore, the greedy heuristic is the more practical optimization routine. A number of factors in our benchmarks contributed to the high performance of the greedy heuristic. To provide a fair comparison with the randomized methods, we chose a moderate range for the generating process, and a reasonably homogeneous pricing structure for the Knapsack and Helipad scenarios, which had the effect that spacing sampling points out evenly over the domain was an approximately optimal strategy. In scenarios with more highly variable or non-linear logistics, or with ranges that are small compared to the domain, both the greedy heuristic and the various design-based methods would be expected to underperform, because they would include points without considering that doing so might exclude a large range of other important points, or that a much cheaper location might supply a nearly equal amount of information, and would therefore space points out farther than is optimal. To provide a conservative comparison, we chose scenarios and generating processes that tended to favor the greedy heuristic. In the future, these methods can be compared under a wider variety of logistical scenarios, such as transect sampling, vehicle routing, and spatio-temporal sampling. They can also be evaluated under further scenarios, such as more severely mis-specified covariance structures. By testing a more diverse set of scenarios, it should be possible to elucidate under what conditions the computational burden of B\&B optimization is beneficial, and when the greedy heuristic can be expected to suffice. 

Notably, the greedy heuristic does not yet have a proven optimality gap for this problem. The B\&B algorithm also did not provide useful optimality gaps, because the LP relaxations largely provided negative variances in the bounding steps. We hypothesize that this is due to the looseness of our Cauchy-Schwartz derived bounds. If tighter bounds can be proven for this problem, we expect that the B\&B algorithm will be able to provide more useful optimality gaps, and will also provide improved solutions more quickly because stronger bounds would improve the efficiency with which branches may be pruned.

We also hope to tackle more complex statistical models using our method in future applications. For example, we hypothesize that non-linear objective functions such as those that would be required for generalized linear mixed models can be approximately accommodated through Taylor series when moderately informative prior information is available or assumed, and that when such prior information is lacking for a small number of variables it may be possible to nevertheless approximately optimize sampling designs for these models by using mixture priors. Such an approach may be particularly advantageous in an adaptive design scenario, when data are available from previous years (e.g. \cite{leach2022recursive}).

Overall, our method formulates a common optimization criterion for spatial surveys -- the uncertainty in the areal mean -- as a MILP so that it can be optimized under complex logistical constraints.  We tested this method using a variety of scenarios and found that under simple logistical constraints it was outperformed by the best design-based methods, but for even moderately complex logistical constraints it outperformed our design-based benchmarks. At present the greedy warmstart algorithm provides comparable results more quickly, but we expect that with the development of tighter bounds, and longer run times for a single model, that the B\&B algorithm will provide better results as well as a dynamically improvable optimality gap.



\section*{Acknowledgements}
We gratefully acknowledge the support of Hans-Erik Andersen and Sarah Ellison for contributing their knowledge of the logistics used in the USFS FIA survey of Alaska, and providing the sampling locations used. 

\vfill

\bibliographystyle{plain}
\bibliography{main}

\begin{thebibliography}{}

\bibitem[Stoica and Babu, 2010]{stoica2010algebraic}
Stoica, P. and Babu, P. (2010).
\newblock Algebraic derivation of elfving theorem on optimal experiment design
  and some connections with sparse estimation.
\newblock {\em IEEE Signal Processing Letters}, 17(8):743--745.

\end{thebibliography}


\begin{thebibliography}{10}

\bibitem{avellar2015multi}
Gustavo~SC Avellar, Guilherme~AS Pereira, Luciano~CA Pimenta, and Paulo Iscold.
\newblock Multi-{UAV} routing for area coverage and remote sensing with minimum
  time.
\newblock {\em Sensors}, 15(11):27783--27803, 2015.

\bibitem{bechtold2005enhanced}
William~A Bechtold and Paul~L Patterson.
\newblock {\em The enhanced forest inventory and analysis program--national
  sampling design and estimation procedures}.
\newblock Number~80. USDA Forest Service, Southern Research Station, 2005.

\bibitem{bellhouse2005systematic}
DR~Bellhouse.
\newblock Systematic sampling methods.
\newblock {\em Encyclopedia of Biostatistics}, 8, 2005.

\bibitem{bertsimas1997introduction}
Dimitris Bertsimas and John~N Tsitsiklis.
\newblock {\em Introduction to linear optimization}, volume~6.
\newblock Athena Scientific Belmont, MA, 1997.

\bibitem{byrd1995limited}
Richard~H Byrd, Peihuang Lu, Jorge Nocedal, and Ciyou Zhu.
\newblock A limited memory algorithm for bound constrained optimization.
\newblock {\em SIAM Journal on Scientific Computing}, 16(5):1190--1208, 1995.

\bibitem{clausen1999branch}
Jens Clausen.
\newblock Branch and bound algorithms-principles and examples.
\newblock {\em Department of Computer Science, University of Copenhagen}, pages
  1--30, 1999.

\bibitem{cplex2020user}
CPLEX.
\newblock {IBM} {ILOG} {CPLEX} {O}ptimization {S}tudio {U}ser’s {M}anual,
  version 12 release 6, 2020.

\bibitem{dantzig1957discrete}
George~B Dantzig.
\newblock Discrete-variable extremum problems.
\newblock {\em Operations research}, 5(2):266--288, 1957.

\bibitem{diaz2018sparse}
Paul Diaz, Alireza Doostan, and Jerrad Hampton.
\newblock Sparse polynomial chaos expansions via compressed sensing and
  {D}-optimal design.
\newblock {\em Computer Methods in Applied Mechanics and Engineering},
  336:640--666, 2018.

\bibitem{dumelle2022comparison}
Michael Dumelle, Matt Higham, Jay~M Ver~Hoef, Anthony~R Olsen, and Lisa Madsen.
\newblock A comparison of design-based and model-based approaches for finite
  population spatial sampling and inference.
\newblock {\em Methods in Ecology and Evolution}, 13(9):2018--2029, 2022.

\bibitem{dupont2021optimal}
Gates Dupont, J~Andrew Royle, Muhammad~Ali Nawaz, and Chris Sutherland.
\newblock Optimal sampling design for spatial capture--recapture.
\newblock {\em Ecology}, 102(3):e03262, 2021.

\bibitem{elfving1952optimum}
Gustav Elfving.
\newblock Optimum allocation in linear regression theory.
\newblock {\em The Annals of Mathematical Statistics}, pages 255--262, 1952.

\bibitem{fischer2021managing}
Samuel~M Fischer, Martina Beck, Leif-Matthias Herborg, and Mark~A Lewis.
\newblock Managing aquatic invasions: optimal locations and operating times for
  watercraft inspection stations.
\newblock {\em Journal of Environmental Management}, 283:111923, 2021.

\bibitem{gurobi}
{Gurobi Optimization, LLC}.
\newblock {Gurobi Optimizer Reference Manual}, 2023.

\bibitem{hughes2008acquiring}
Robert~M Hughes and David~V Peck.
\newblock Acquiring data for large aquatic resource surveys: the art of
  compromise among science, logistics, and reality.
\newblock {\em Journal of the North American Benthological Society},
  27(4):837--859, 2008.

\bibitem{kermorvant2019optimizing}
Claire Kermorvant, Nathalie Caill-Milly, No{\"e}lle Bru, and Frank d'Amico.
\newblock Optimizing cost-efficiency of long term monitoring programs by using
  spatially balanced sampling designs: The case of manila clams in {A}rcachon
  {B}ay.
\newblock {\em Ecological Informatics}, 49:32--39, 2019.

\bibitem{krause2008near}
Andreas Krause, Ajit Singh, and Carlos Guestrin.
\newblock Near-optimal sensor placements in gaussian processes: Theory,
  efficient algorithms and empirical studies.
\newblock {\em Journal of Machine Learning Research}, 9(Feb):235--284, 2008.

\bibitem{kwak2022sensor}
Dongho Kwak, Joonsoo Jeong, Yongbeom Shin, Nagyeong Lee, and Dongil Shin.
\newblock Sensor placement optimization for fenceline monitoring of toxic gases
  considering spatiotemporal risk of the plant-urban interface.
\newblock {\em Journal of the Taiwan Institute of Chemical Engineers},
  130:103858, 2022.

\bibitem{land2010automatic}
Ailsa~H Land and Alison~G Doig.
\newblock An automatic method for solving discrete programming problems.
\newblock In {\em 50 Years of Integer Programming 1958-2008}, pages 105--132.
  Springer, 2010.

\bibitem{leach2022recursive}
Clinton~B Leach, Perry~J Williams, Joseph~M Eisaguirre, Jamie~N Womble,
  Michael~R Bower, and Mevin~B Hooten.
\newblock Recursive bayesian computation facilitates adaptive optimal design in
  ecological studies.
\newblock {\em Ecology}, 103(2):e03573, 2022.

\bibitem{lee2009non}
Jon Lee, Vahab~S Mirrokni, Viswanath Nagarajan, and Maxim Sviridenko.
\newblock Non-monotone submodular maximization under matroid and knapsack
  constraints.
\newblock In {\em Proceedings of the Forty-First Annual ACM Symposium on Theory
  of Computing}, pages 323--332, 2009.

\bibitem{lindgren2011explicit}
Finn Lindgren, H{\aa}vard Rue, and Johan Lindstr{\"o}m.
\newblock An explicit link between gaussian fields and gaussian markov random
  fields: The stochastic partial differential equation approach.
\newblock {\em Journal of the Royal Statistical Society: Series B (Statistical
  Methodology)}, 73(4):423--498, 2011.

\bibitem{lohr2019sampling}
Sharon~L Lohr.
\newblock {\em Sampling: Design and Analysis: Design and Analysis}.
\newblock Chapman and Hall/CRC, 2019.

\bibitem{pukelsheim2006optimal}
Friedrich Pukelsheim.
\newblock {\em Optimal Design of Experiments}.
\newblock SIAM, 2006.

\bibitem{rao1990history}
JNK Rao and DR~Bellhouse.
\newblock History and development of the theoretical foundations of survey
  based estimation and analysis.
\newblock {\em Survey Methodology}, 16(1):3--29, 1990.

\bibitem{robertson2013bas}
BL~Robertson, JA~Brown, Trent McDonald, and Peter Jaksons.
\newblock {BAS}: Balanced acceptance sampling of natural resources.
\newblock {\em Biometrics}, 69(3):776--784, 2013.

\bibitem{sagnol2010optimal}
Guillaume Sagnol, St{\'e}phane Gaubert, and Mustapha Bouhtou.
\newblock Optimal monitoring in large networks by successive {C}-optimal
  designs.
\newblock In {\em 2010 22nd International Teletraffic Congress (lTC 22)}, pages
  1--8. IEEE, 2010.

\bibitem{stevens2004spatially}
Don~L Stevens~Jr and Anthony~R Olsen.
\newblock Spatially balanced sampling of natural resources.
\newblock {\em Journal of the American Statistical Association},
  99(465):262--278, 2004.

\bibitem{stoica2010algebraic}
Petre Stoica and Prabhu Babu.
\newblock Algebraic derivation of {E}lfving theorem on optimal experiment
  design and some connections with sparse estimation.
\newblock {\em IEEE Signal Processing Letters}, 17(8):743--745, 2010.

\bibitem{tobler1970computer}
Waldo~R Tobler.
\newblock A computer movie simulating urban growth in the {D}etroit region.
\newblock {\em Economic Geography}, 46(sup1):234--240, 1970.

\bibitem{van1999constrained}
Jan~Willem Van~Groenigen, W~Siderius, and A~Stein.
\newblock Constrained optimisation of soil sampling for minimisation of the
  kriging variance.
\newblock {\em Geoderma}, 87(3-4):239--259, 1999.

\bibitem{wang2012review}
Jin-Feng Wang, A~Stein, Bin-Bo Gao, and Yong Ge.
\newblock A review of spatial sampling.
\newblock {\em Spatial Statistics}, 2:1--14, 2012.

\bibitem{white1992cartographic}
Denis White, Jon~A Kimerling, and Scott~W Overton.
\newblock Cartographic and geometric components of a global sampling design for
  environmental monitoring.
\newblock {\em Cartography and Geographic Information Systems}, 19(1):5--22,
  1992.

\bibitem{williams2013model}
H~Paul Williams.
\newblock {\em Model Building in Mathematical Programming}.
\newblock John Wiley \& Sons, 2013.

\bibitem{wynn1970sequential}
Henry~P. Wynn.
\newblock The sequential generation of {$D$}-optimum experimental designs.
\newblock {\em The Annals of Mathematical Statistics}, 41(5):1655 -- 1664,
  1970.

\bibitem{yilmaz2008path}
Namik~Kemal Yilmaz, Constantinos Evangelinos, Pierre~FJ Lermusiaux, and
  Nicholas~M Patrikalakis.
\newblock Path planning of autonomous underwater vehicles for adaptive sampling
  using mixed integer linear programming.
\newblock {\em IEEE Journal of Oceanic Engineering}, 33(4):522--537, 2008.

\end{thebibliography}

\end{document}


\def\spacingset#1{\renewcommand{\baselinestretch}%
{#1}\small\normalsize} \spacingset{1}


\if1\blind
{
  \title{\bf Appendix to: Optimal Sampling Design Under Logistical Constraints with Mixed Integer Programming}
  \author{Connie Okasaki\thanks{
    This material is based upon work supported by the National Science Foundation Graduate Research Fellowship under Grant No. DGE-1762114}\hspace{.2cm}\\
    Quantitative Ecology and Resource Management Program, U of Washington,\\
    S\'{a}ndor Toth \\
    School of Environmental and Forestry Sciences, U of Washington, \\
    and \\
    Andrew M. Berdahl\thanks{AMB was supported by the H. Mason Keeler Endowed Professorship in Sports Fisheries Management.} \\
    School of Aquatic and Fisheries Sciences, U of Washington}
  \maketitle
} \fi

\if0\blind
{
  \bigskip
  \bigskip
  \bigskip
  \begin{center}
    {\LARGE\bf Title}
\end{center}
  \medskip
} \fi

\noindent%

\pagebreak 

\section{Optimal Sampling Design with Linear Regression}\label{app:C1}

Suppose that there exist known covariates $\bm{X}$ within the region of interest, and that we wish to control for the effects of these covariates on our estimates. This can be done within a hierarchical Bayesian framework without losing any of the convenient properties derived above. Specifically, by placing a normal prior (equivalent to ridge regression) upon the regression coefficients, we retain an entirely multivariate normal conditional distribution after observing the data:
\begin{align*}
\bm{d} & = \bm{A}^T\bm{u} + \bm{\eps}, & \bm{\eps} & \sim \mathcal{N}(0,\sigma^2_\eps \bm{I}_n), \\
\bm{u} & = \bm{X\beta} + \bm{\eta}, & \bm{\eta} & \sim \mathcal{N}(0,\bm{Q}), \\
&& \bm{\beta} & \sim \mathcal{N}(0,\sigma^2_\beta \bm{I}_k).
\end{align*}
where $k$ is the number of covariates and $n$ the number of data points. This gives us the following conditional distribution after observing $\bm{d}$
\begin{align*}
[\bm{\eta},\bm{\beta}|\bm{d}] 
& = [\bm{d}|\bm{\eta},\bm{\beta}][\bm{\eta}][\bm{\beta}] \\
& = \mathcal{N}(\bm{X\beta} + \bm{\eta},\sigma^2_\eps \bm{I}_n)\mathcal{N}(0,\bm{Q})\mathcal{N}(0,\sigma^2_\beta \bm{I}_k).
\end{align*}
This distribution has a precision matrix of
\begin{align*}
\bm{Q}' & = \begin{bmatrix} \bm{Q} + \frac{1}{\sigma^2_\eps}\bm{A}\bm{A}^T & \frac{1}{\sigma^2_\eps}\bm{A}\bm{A}^T\bm{X} \\ \frac{1}{\sigma^2_\eps}\bm{X}^T\bm{A}\bm{A}^T & \frac{1}{\sigma^2_\beta}\bm{I}_k + \frac{1}{\sigma^2_\eps}\bm{X}^T\bm{A}\bm{A}^T\bm{X}\end{bmatrix}.
\end{align*}
This matrix can be substituted for the matrix $\bm{Q}$ in the objective function and the remainder of the model building proceeds analogously. The vector $\bm{v}$ will have terms for each of the covariates corresponding to the integrated value of that covariate across the spatial area. The terms involving $\bm{X}$ will often be dense, but so long as the number of parameters $k$ remains small the full matrix should remain relatively sparse, and optimization should remain computationally feasible. 

\section{Discrete Uncertainty in the Precision Matrix}\label{app:C2}

There are at least two notions of uncertainty that one might apply to the precision matrix. The first is a discrete distribution for $\bm{Q}$ in which several fixed options $\bm{Q}_i$ are weighted with probabilities $p_i$. Such a situation may arise due to uncertainty in the functional form of the covariance function, or as an approximation to a continuous distribution for $\bm{Q}$. In this situation, our objective function becomes probabilistic, with mean given by:
\[
\sum_i p_i \bm{v}^T(\bm{Q}_i + \sum x_i \bm{a}_i\bm{a}_i^T)^{-1}\bm{v}.
\]
This objective function may still be used within our MILP, as follows:

\begin{mini}
{\bm{x},\bm{y}_j,\bm{z}_j}{\sum p_j(\bm{v}^T\bm{y_j}),}
{\label{eq:obj-example}}{}
\addConstraint{\bm{Q}_j\bm{y}_j + \frac{1}{\sigma^2}\bm{A}^T\bm{z}_j}{= \bm{v}}
\addConstraint{\bm{x}_i*\bm{a}_i^T\bm{y}_j}{= \bm{z}_jm,} \\
\addConstraint{\bm{x}_i}{\in \{0,1\}.}
\end{mini}

We know of no way to represent continuous uncertainty in $\bm{Q}$ will maintaining model linearity. 

\section{Using the Covariance Matrix}\label{app:C3}
Some spatial statistical methods use sparse versions of the covariance matrix instead. These remain consistent with our approach. To utilize these methods, all one needs do is modify the constraint
\[
\bm{Q}\bm{y} + \frac{1}{\sigma^2}\bm{A}^T\bm{z} = \bm{v}
\]
to
\[
\bm{y} + \frac{1}{\sigma^2}\bm{\Sigma}\bm{A}^T\bm{z} = \bm{\Sigma}\bm{v}.
\]
Both the matrix $\bm{\Sigma A}^T$ and the vector $\bm{\Sigma v}$ may be precalculated.

\section{Design-Based Knapsack Sampling}\label{app:C4}

\subsection{Without Spatial Covariance}
Sampling under knapsack constraints can be accomplished using methods similar to those in \cite{stoica2010algebraic}. This method may accommodate regression models, as well as spatial balancing, but may not accommodate explicit accounting for spatial covariance without additional modification. Using Stoica and Babu's notation, assume that we have a set of $K$ regression vectors $\bm{a}_k$, each of which is sampled in proportion equal to $p_k$ of the total number of sites (proportions are used by Stoica, but we can just as easily imagine using these ``proportions'' as weights). In this case, the expected Fisher information matrix is given by Stoica and Babu

\[
E[\bm{R}] = \sum_{k=1}^K p_k\bm{a}_k\bm{a}_k^T = \bm{A}^T\bm{P}\bm{A},
\]

where
\begin{align*}
    \bm{A}^T & = \begin{bmatrix} \bm{a}_1 & ... & \bm{a}_K\end{bmatrix}, \\
    \bm{P} & = \begin{bmatrix} p_1 & 0 & ... & 0 \\ 0 & p_2 & ... & 0, \\ \vdots & \vdots & \ddots & \vdots \\
    0 & ... & ... & p_K\end{bmatrix} \\
    \mbox{and} & \sum_{k=1}^K p_k = 1.
\end{align*}

We wish to modify the final constraint to a more general knapsack type constraint $\sum_{k=1}^K c_k\tilde{\bm{P}}_k = B$, so that the expected cost is equal to a given budget $B$. To accomplish this, we may set $\tilde{\bm{P}}_k = \frac{B}{c_k}p_k$ so that
\[
\sum_{k=1}^K c_k\tilde{\bm{P}}_k = B\sum_{k=1}^K p_k = B,
\]
and 
\[
\bm{R} = \sum_{k=1}^K \tilde{\bm{P}}_k\bm{a}_k\bm{a}_k^T = \sum_{k=1}^K p_k \left(\sqrt{\frac{B}{c_k}}\bm{a}_k\right)\left(\sqrt{\frac{B}{c_k}}\bm{a}_k\right)^T.
\]
Therefore whatever the minimum $\bm{R}$ that can be achieved, either parameterization will be able to achieve it. One may therefore leverage the homotopy method suggested by Stoica and Babu to solve the modified problem by transforming the regression vectors to
\[
\sqrt{\frac{B}{c_k}}\bm{a}_k
\]
and then transforming the resulting $p_k$ to obtain the knapsack constrained sampling weights. These weights may be plugged into an SBRS procedure (e.g. GRTS) to obtain a spatially balanced, knapsack constrained sampling design. 

\subsection{With Spatial Covariance}
In the event that one wishes to collect a design-based sample with respect to a particular spatial covariance function, this may be accomplished with some additional modification to the methods. Let us suppose we wish to minimize 
\[
\bm{v}^T(\bm{Q} +\sum_{i}p_i\bm{a}_i\bm{a}_i^T)^{-1}\bm{v} = \bm{v}^T(\bm{Q}+\bm{APA}^T)^{-1}\bm{v},
\]
the conditional variance of a particular linear combination $w$ of points (e.g. the total kriging variance in space). This equation strongly resembles that in the Elfving theorem, so we will follow an algebraic derivation to arrive at a similar result. First we will introduce the augmented problem
\begin{mini*}|l|{\bm{p},\bm{c},\bm{d}}{h = \bm{c}^T\bm{P}^+\bm{c} + \bm{d}^T\bm{d},}{}{}
\addConstraint{\bm{A}^T\bm{c} + \bm{Q}^{1/2}\bm{d}}{=\bm{v}}{}{}
\addConstraint{\sum p_i}{=1}{}{}
\addConstraint{c_i}{=0}{\iff}{p_i=0.}
\end{mini*}
We propose that, for fixed $p$ this augmented minimization problem has solution 
\begin{align*}
\bm{c}_0 & = \bm{PA}^T(\bm{Q}+\bm{APA}^T)^{-1}\bm{v}, & \bm{d}_0 & = \bm{Q}^{1/2}(\bm{Q}+\bm{APA}^T)^{-1}\bm{v},
\end{align*}
and thus objective function $h_0 = \bm{v}^T(\bm{Q}+\bm{APA}^T)^{-1}\bm{v}$ as above. If we can verify this claim then the augmented problem will have the same minimum as the simpler problem above. We now show that this is correct:
\begin{align*}
h \geq h_0 
& \iff \begin{bmatrix}\bm{c}_0^T\bm{P}^+\bm{c}_0 + \bm{d}_0^T\bm{d}_0 & \bm{v}^T \\ \bm{v} & \bm{Q}+\bm{A}^T\bm{PA}\end{bmatrix} \geq 0, \\
& \iff \begin{bmatrix}\tilde{\bm{c}}_0^T\tilde{\bm{P}}^+\tilde{\bm{c}}_0 + \bm{d}_0^T\bm{d}_0 & \tilde{\bm{c}}_0^T\tilde{\bm{A}} + \bm{d}_0^T\bm{Q}^{1/2} \\ \tilde{\bm{A}}^T\tilde{\bm{c}}_0 + \bm{Q}^{1/2}\bm{d}_0 & \bm{Q}+\tilde{\bm{A}}^T\tilde{\bm{P}}\tilde{\bm{A}}\end{bmatrix} \geq 0, \\
& \iff \begin{bmatrix}\tilde{\bm{c}}_0^T& 0 \\0 & \tilde{\bm{A}}^T\end{bmatrix}\begin{bmatrix}\tilde{\bm{P}}^{-1/2}\\\tilde{\bm{P}}^{1/2}\end{bmatrix}\begin{bmatrix}\tilde{\bm{P}}^{-1/2}&\tilde{\bm{P}}^{1/2}\end{bmatrix}\begin{bmatrix}\tilde{\bm{c}}_0 & 0 \\0 & \tilde{\bm{A}}\end{bmatrix} + \begin{bmatrix}\bm{d}_0^T \\ \bm{Q}^{1/2} \end{bmatrix}\begin{bmatrix} \bm{d}_0 & \bm{Q}^{1/2} \end{bmatrix} \geq 0,
\end{align*}
where $\tilde{\bm{c}}$ and $\tilde{\bm{P}}$ and $\tilde{\bm{A}}$ all correspond to matrices with the zero elements from $\bm{c}$ and $\bm{P}$ removed. 

This last expression is the sum of two PSD matrices, so we find that indeed, $h_0$ is the minimum. Now let us suppose that $\bm{c}$ is fixed. Because $\bm{Q}$ is nonsingular, the equality constraints gives us a solution for $\bm{d}$:
\begin{align*}
\bm{d} & = \bm{Q}^{-1/2}(\bm{v}-\bm{A}^T\bm{c}).
\end{align*}
Then we find that $P$ is minimized just as if $\bm{d}$ did not exist (following \citealt{stoica2010algebraic}) at
\begin{align*}
p_i = \frac{|c_i|}{\sum_j |c_j|}
\end{align*}
This can be argued by Cauchy-Schwartz, because this choice of $p_i$ achieves $\bm{c}^T\bm{P}^+\bm{c} = \left(\sum_i|c_i|\right)^2$, but
\begin{align*}
\left(\sum_i |c_i|\right)^2 = \left(\sum_{i} \frac{|c_i|}{\sqrt{p_i}}\sqrt{p_i}\right)^2 \leq \left(\sum_i \frac{c_i^2}{p_i}\right)\left(\sum_i p_i\right) = \left(\sum_i \frac{c_i^2}{p_i}\right)
\end{align*}

Now because $\bm{c}$ is not fixed, we re-express $\bm{p}$ and $\bm{d}$ in terms of $\bm{c}$ and our minimization problem becomes becomes:
\begin{mini*}|l|{c}{\left(\sum_i |c_i|\right)^2 + (\bm{v}-\bm{A}^T\bm{c})^T\bm{Q}^{-1}(\bm{v}-\bm{A}^T\bm{c}).}{}{}
\end{mini*}
This is a linearly constrained convex quadratic program of the form
\begin{mini*}|l|{c}{\bm{z}^T\bm{z} + (\bm{v}-\bm{A}^T\bm{c})^T\bm{Q}^{-1}(\bm{v}-\bm{A}^T\bm{c}),}{}{}
\addConstraint{z_i}{\geq c_i}{}{}
\addConstraint{z_i}{\geq -c_i.}{}{}
\end{mini*}{}
This problem may lose the sparsifying effect from the $\ell_1$ minimization that may be used in the absence of an explicit covariance term, due to the now-quadratic objective function. Similar to the above, we may allow variable costs ($\sum q_ip_i = B$) by normalizing with respect to budget ($\sum \frac{q_i}{B}p_i = 1$) and then subsuming cost fraction into observations. That is:
\begin{mini*}|l|{p}{\bm{v}^T(\bm{Q}+\sum p_i\bm{a}_i\bm{a}_i^T)^{-1}w,}{}{}
\addConstraint{\sum q_ip_i}{\leq B,}{}{}
\end{mini*}
is equivalent to
\begin{mini*}|l|{p}{\bm{v}^T\left(\bm{Q}+\sum p_i\left(\sqrt{\frac{B}{q_i}}\bm{a}_i\right)\left(\sqrt{\frac{B}{q_i}}\bm{a}_i\right)^T\right)^{-1}\bm{v},}{}{}
\addConstraint{\sum p_i}{\leq 1.}{}{}
\end{mini*}
Covariates may be included in the same way as in our main MILP model.

\pagebreak

\bibliographystyle{apalike}
\bibliography{../0main}